\documentclass[aps,pre,showpacs,twocolumn,superscriptaddress,showkeys]{revtex4-1}
\usepackage{graphicx}

\begin{document}
\title{Weyl asymptotics: From closed to open systems}

\author{A.~Potzuweit}
\affiliation{Fachbereich Physik, Philipps-Universit\"{a}t Marburg, Renthof 5,
35032 Marburg, Germany}
\author{T.~Weich}
\affiliation{Fachbereich Physik, Philipps-Universit\"{a}t Marburg, Renthof 5,
35032 Marburg, Germany}
\affiliation{Fachbereich Mathematik, Philipps-Universit\"{a}t Marburg, Hans-Meerwein-Stra{\ss}e,
35032 Marburg, Germany}
\author{S.~Barkhofen}
\affiliation{Fachbereich Physik, Philipps-Universit\"{a}t Marburg, Renthof 5,
35032 Marburg, Germany}
\author{U.~Kuhl}
\affiliation{Laboratoire de Physique de la Mati\`{e}re Condens\'{e}e, CNRS UMR 7336, Universit\'{e} de Nice
Sophia-Antipolis, F-06108 Nice, France} \email{ulrich.kuhl@unice.fr}
\affiliation{Fachbereich Physik, Philipps-Universit\"{a}t Marburg, Renthof 5,
35032 Marburg, Germany}
\author{H.-J.~St\"{o}ckmann}
\affiliation{Fachbereich Physik, Philipps-Universit\"{a}t Marburg, Renthof 5,
35032 Marburg, Germany}
\author{M.~Zworski}
\affiliation{Department of Mathematics, University of California, Berkeley, California 94720, USA}

\date{\today}
\begin{abstract}
We present microwave experiments on the symmetry reduced five-disk billiard studying the transition from a closed to an open system. The measured microwave reflection signal is analyzed by means of the harmonic inversion and the counting function of the resulting resonances is studied. For the closed system this counting function shows the Weyl asymptotic with a leading exponent equal to 2. By opening the system successively this exponent decreases smoothly to an non-integer value. For the open systems the extraction of resonances by the harmonic inversion becomes more challenging and the arising difficulties are discussed. The results can be interpreted as a first experimental indication for the fractal Weyl conjecture for resonances.
\end{abstract}

\keywords{Weyl law, fractal repeller, n-disk system, harmonic inversion}
\pacs{05.45.Mt, 03.65.Nk, 42.25.Bs,25.70.Ef}

\maketitle

\section{Introduction}
\label{sec:Introduction}

In this year we celebrate the 100th anniversary of the Weyl law, which in its earliest version gives the leading term in the asymptotic description of the counting function of the Dirichlet or Neumann Laplacien on a bounded domain in the Euclidean space \cite{wey12a}. Weyl's original intention was to justify the derivation of the Rayleigh-Jeans law \cite{jea1905} for an arbitrary domain, not just boxes. Following works by many mathematicians, among them Courant, Hilbert, Avakumovic, and Levitan, in 1968 H\"{o}rmander obtained a very general Weyl law for elliptic operators \cite{hoer68}. The improved remainders under dynamical assumptions were obtained by Duistermaat-Guillemin on manifolds without boundary \cite{dui75} and by Ivrii, who proved the Weyl conjecture on the second term in asymptotics [see (\ref{eq:WeylClosed}) below] \cite{ivr80a,ivr82a}. In physics literature, higher order terms for smoothed out counting function were introduced already in the 1970s by Balian and Bloch in the context of electromagnetic eigenmodes in cavities with perfectly conducting smooth walls \cite{bal70}. These results were experimentally verified in three and two dimensional microwave cavities \cite{deu95,grae92}. It is intriguing that for typical ``closed'' cavities the Weyl law agrees well even down to the ground state. In more recent developments the Weyl formula has been extended to take into account bouncing ball fluctuations \cite{grae92,sie93}, fractal boundaries \cite{ber79b,bro86}, and ray-splitting \cite{pra96b}.

As soon as a wave-mechanical system is open the eigenvalues turn into resonances. The study of eigenvalues, resonances, and quasibound states has a long tradition in theoretical, numerical, and experimental chaotic scattering (see, for instance, \cite{gas89a} and references therein). One argument for the importance of open systems is the fact that for a measurement it is inevitable to allow interaction with the outside of the system which effectively makes it open. If the system is opened only weakly, the Weyl formula is a very good approximation, as one can already see from the microwave experiments mentioned before. However, if the system is strongly coupled to the environment the resonances cannot be related to individual eigenvalues of the closed systems. If the number of attached channels is finite the framework of random matrix theory (RMT) using non-Hermitian Hamiltonians predicts a separation of nearby resonances \cite{joh91,haa92,pol12}. By varying the coupling resonance trapping effects show up, that is, by
increasing coupling many resonances may become sharper and only a few become much broader \cite{oko03,rot09,per00}. All these effects increase the difficulty of counting resonances compared to the counting of eigenvalues.

If one is interested in the description of the resonances in the semiclassical limit the number of channels diverges and random matrix description breaks down. In this case many quantum mechanical quantities are determined by quantities of the underlying classical dynamics.  For example, Gaspard and Rice \cite{gas89a} derived a semiclassical lower bound on the resonance width and Sridhar observed experimentally that the quantum escape rate for $n$-disk systems coincides with half of the classical escape rate \cite{lu99}. For open dynamical systems with a fractal repeller, Sj\"{o}strand showed that the counting function is polynomially bounded with the power given by the box dimension of the classical repeller \cite{sjo90}.
In systems with only two degrees of freedom, such as considered here, this corresponds to the Hausdorff dimension. The counting function will be introduced in Sec.~\ref{sec:CF}. Numerical investigations in \cite{lin02b} and \cite{lu03} lead to a conjecture that this bound is, in fact, optimal. Numerous numerical and theoretical studies in physics and mathematics followed \cite{zwo99b,gui04,sch04e,non05,kea06b,sjo07}. In particular, the Weyl law has been investigated numerically on maps, the kicked rotator, or the three-disk billiard \cite{lu03,sch04e,non05,kea06b,kop10}. The fractal dimension of the chaotic set also occurs in systems with mixed phase space \cite{kop10,erm10,erm11,ebe10,ped09,ped12,ram09,str04,wie08a,sch09,chr07}. Until now there exists no experimental verification for this law as it is very challenging to extract resonances of strongly open systems in the semiclassical regime. The most recent mathematical advances include fractal upper bounds for several convex bodies \cite{arXnon1} and for
arbitrary manifolds with hyperbolic ends \cite{arXdat1}.

The paradigmatic physical example of an open system with fractal repeller is the $n$-disk system which has been introduced in the context of resonances by Ikawa in mathematics \cite{ika88} and by Gaspard and Rice \cite{gas89b,gas89a} and Cvitanovi{\'c} and Eckhardt \cite{cvi89} in physics. Theoretically it has been studied in the classical, semiclassical, and quantum mechanical regime and there have also been an experimental study on the spectral autocorrelation \cite{pan00a,lu00}. In this article we want to investigate the counting function of a symmetry reduced $n$-disk system experimentally. We realize the ``quantum'' $n$-disk system by a microwave cavity and use the well established equivalence between Helmholtz and Schr\"odinger equation \cite{stoe99} which makes it possible to measure wave characteristics of quantum single-particle systems with a table-top microwave experiment. As we are interested in the transition from a closed to an open system, we chose the five-disk system, as for this system the completely closed system is already sufficiently large for performing experiments. Experimentally, we are restricted to a finite frequency range leading to about 150 resonances. We start from a closed system and by changing a parameter the system opening increases (see Sec.~\ref{sec:ExpSetup}). For small openings the system only couples via few channels, whereas for large opening a ``semiclassical'' coupling will be realized. The resonances are extracted by the method of harmonic inversion (see Sec.~\ref{sec:HI}) and in Sec.~\ref{sec:ExpResults} we present the experimental findings. Concluding remarks are given in Sec.~\ref{sec:Conclusion}.


\section{Eigenvalue and resonance counting functions}
\label{sec:CF}
As we are dealing with a two dimensional system with hard wall potentials we will restrict the discussion of the counting function to this case. If we consider a two dimensional quantum mechanical closed system with hard wall potential, then it is described by the Laplace operator with Dirichlet boundary conditions which has a discrete real spectrum $\Delta \psi_n = -k_n^2\psi_n$. The system's counting function $N(k)$ is then defined as
\begin{equation} \label{eq:DefNkReal}
 N(k):=\#\{k_n: k_n\leq k\}
\end{equation}
If the area of the system is denoted by $A$ and its circumference by $U$ the Weyl law predicts, that the counting function is given by
\begin{equation} \label{eq:WeylClosed}
 N_\mathrm{Weyl}(k) = \frac{ 1 }{ 4 \pi } A k^2\pm \frac{ 1} { 4 \pi } U k+c
\end{equation}
where $\pm$ depends on the boundary condition: $+$ for Neumann and $-$ for Dirichlet boundary conditions. The constant $c$ is defined by the curvature and the corners in the system \cite{bal76}. In general the leading term is proportional to $k^d$, where $d$ is the dimensionality of the system. The Weyl law (\ref{eq:WeylClosed}) corresponds to the counting function (\ref{eq:DefNkReal}) only after appropriate smoothing of the counting function.

By coupling the system to the environment, eigenvalues and eigenfunctions turn into resonances and scattering states. One possible description is via a non-Hermitian effective Hamiltonian \cite{oko03,rot09}. If the coupling is performed only via a few weakly coupled channels all resonances will only acquire a small imaginary part. Thus, counting can be performed for the real part of the resonances and the number of resonances will satisfy the Weyl law (\ref{eq:WeylClosed}). To keep the coupling small and the number of channels constant in the semiclassical limit is not realistic.

Let us first assume that the number of channels $M$ stays constant but the coupling of all channels increases. This corresponds to a RMT where the restriction of Hermiticity is dropped. Its properties have been investigated and for strong coupling a separation of resonance by their imaginary part is given \cite{haa92}. If the system contains $N>M$ resonances then one finds $M$ broad resonances, that is, resonances with large imaginary parts, and $N-M$ resonances with small imaginary parts. Now one has to define whether all resonances are to be counted or only resonances up to a certain imaginary part $C$. The counting statistics of resonances is typically defined by
\begin{equation}\label{eq:DefCounting}
 N(k) := \#\{ \tilde{k}_n: \mathrm{Im}\,\tilde{k}_n > -C, \mathrm{Re}\,\tilde{k}_n \leq k \}
\end{equation}
where $C$ is a fixed finite positive constant, which should not be too small. Thus depending on $C$ one would either count $N$ or $N-M$ resonances. This effect is not restricted to RMT models but holds for any wave system strongly coupled to the environment. It has been phrased as resonance trapping in the framework of the effective Hamiltonian theory (for details, see \cite{rot09} and references therein). It has also been observed experimentally in microwave cavities with variable coupling \cite{per00,stoe02c}. If the maximal wave number $k_\mathrm{max}$ is not too large, this might be observable in the counting function as well (see, e.g., Fig.~5 of \cite{stoe02c}). If waveguides with width $d$ are attached, each of them supports $M_w=d/(\lambda/2)=dk/\pi$ modes, where $\lambda$ is the wavelength. Thus, the number of waveguide modes and therefore also the number of channels increases linearly in the semiclassical limit, whereas the number of eigenvalues of the closed cavity increases quadratically. This would
still lead to a dominating Weyl term of $k^2$. Another possibility is to couple as many channels as eigenvalues (or more) to the system. In the framework of RMT this is related to the Ginibre ensemble \cite{gin65,haa01b}.

Until now we have neglected any additional variation of the real part of the resonances that might be induced by changing the coupling as well as the internal classical dynamics. The structure of this internal dynamics will induce special phase space structures leading to deviations from RMT predictions. The classical phase space of open chaotic systems is characterized by the forward and backward trapped sets. The repeller is the intersection of the fractal sets of trajectories that stay trapped forever in the future and in the past. In the semiclassical limit the wave functions of the long-lived resonances, that is, resonances with small imaginary parts, will localize on the trapped set \cite{erm09,ped12}. Thus, they will avoid the coupling to the channels which is in correspondence with the resonance trapping effect.

As indicated in the Introduction the mathematical and numerical works suggest that the counting statistics in case of open systems with a classical fractal repeller correspond to a fractal Weyl law:
\begin{equation}\label{eq:FractalWeyl}
 N(k) \sim k^{1+d_H},
\end{equation}
where $d_H$ is the reduced fractal dimension of the repeller. We stress that the rigorous results so far prove only an upper bound of this form, or finer bounds in smaller intervals \cite{sjo07,arXnon1} and for arbitrary manifolds with hyperbolic ends \cite{arXdat1}. The numerical papers on maps \cite{non05}, open cavities \cite{lu03}, and microdisk lasers \cite{wie08a} showed the asymptotic behavior.

In this article we investigate experimentally a transition from a closed system to an open five-disk system. Experimentally, one is typically restricted to the wave number $k$ range to $0 \le k \le k_\mathrm{max}$, corresponding to approximate 150 resonances in our case. The billiard of interest consists of two straight walls inducing Dirichlet boundary conditions with an angle of $\theta=36^\circ$. A half circle is attached to only one wall with radius $a$ and distance $d$ from the corner. The position $d$ can be varied but angle and radius of the disk are fixed. In the following section we describe the experimental setup in detail.


\section{Experimental Setup}
\label{sec:ExpSetup}

\begin{figure}
\includegraphics[width=.8\columnwidth]{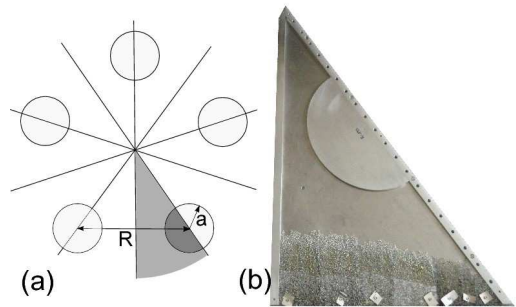}
\caption{\label{fig:setup}
(a) A sketch of the five-disk scattering system is given. The gray shaded part shows the fundamental domain of the symmetry reduction. The radius of the disks is denoted by $a$ and their distance by $R$. (b) A photograph of the experimental setup is presented.
}
\end{figure}

In the experiments we use a flat microwave resonator with a single wire antenna inserted through a hole in the top plate [see Fig.~\ref{fig:setup}(b)]. The baseplate is an aluminum triangle whose largest side is $100$\,cm long. Two side walls with a height of $6$\,mm are set atop; the angle between them is $36^\circ$. Thus the resonator forms the fundamental domain of a five disk system with metallic sidewalls, which act as mirrors reducing the symmetry of the system. The metallic walls induce Dirichlet boundary conditions for the electric field of the TM$_0$ mode, thereby restricting the measurements to a single representation $A_{1u}$ of the underlying symmetry group $C_{5v}$ [see sketch in Fig.~\ref{fig:setup}(a)]. The third side is left open, but additionally covered with a wide strip of microwave absorber to avoid reflections at the open end. Along the longest triangle side we move a half-disk inset of radius $a=19.5\,$cm, with the same height as the side walls, in steps of 10\,
mm. The cover plate is not shown in Fig.~\ref{fig:setup}. An antenna is inserted through a hole in the upper corner of the cover. The radius of the antenna ($r$=0.7\,mm) is much smaller than the wavelength and the antenna is sufficiently short not to touch the bottom plate. The reflection coefficient is measured with a vector-network-analyzer (VNA), revealing the complex $S$ matrix. The height of the cavity $h=6$\,mm (along the vertical $z$ direction) leads to a cutoff frequency of 25\,GHz, below which the $x$- and $y$ component of the electro-magnetic field must vanish and only $E_z$ remains. These electromagnetic modes with vanishing $E_x$ and $E_y$ components are also called TM$_0$ modes. By limiting the frequency range to be analyzed to 24\,GHz, we thus make sure that only the TM$_0$ mode can propagate and the cavity may be considered as two dimensional. The Helmholtz equation describing the electromagnetic wave can then be written as
\begin{equation}\label{eq:Helmholtz}
 \Delta_{xy} \psi_n = -k_n^2\psi_n
\end{equation}
where the wave function $\psi_n = E_z$ and the wavenumber $k_n = 2\pi\nu_n/c$ with $\nu_n$ being the eigenfrequency of the electromagnetic wave. For Dirichlet boundary conditions ($E_z$ must vanish at the side walls) there exists a one-to-one correspondence between wave mechanics and quantum mechanics, that is, between the time independent Helmholtz and Schr\"{o}dinger equation \cite{stoe99}.

The disk positions are characterized by a distance-to-radius parameter $R/a$, the ratio between the distance $R$ of two disks in the full system and the disk radius $a$ (see Fig.~\ref{fig:setup}). The antenna is placed in between the acute angle of the aluminum walls and the movable disk inset. The position is chosen such, that the antenna is not to close to any wall even in the closed case for $R/a =2$. Thus, always a coupling to the interior of the scattering system is guaranteed, where the long-living states live. While increasing the ratio $R/a$ from 2 to 3.9 the fractal dimension of the underlying classical repeller changes and accordingly should the exponent of the counting function. The whole setup is covered by a second aluminum plate that is firmly pressed onto the billiard, leaving no gap between walls and cover plate.


\section{Harmonic Inversion}
\label{sec:HI}

\begin{figure}
\includegraphics[width= 0.45\textwidth]{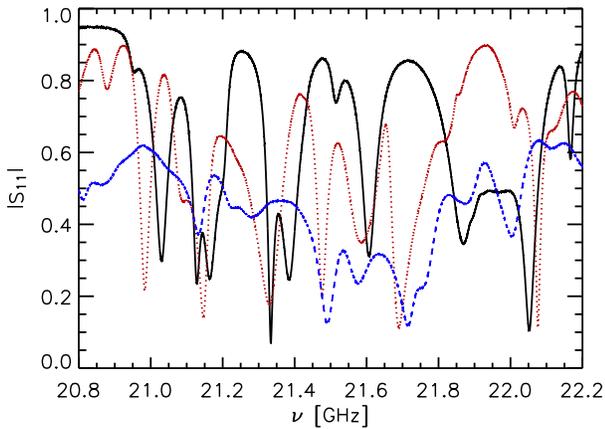}
\caption{\label{fig:Spec} (Color online)
The plot shows measured reflection spectra for the closed system $R/a=2$ (black solid line),
a system in the transition region with $R/a=2.25$ (red dotted line), and an open system with $R/a = 3.83$ (blue dashed line).}
\end{figure}

Under the assumption that the antenna couples point-like to the system the measured reflection signal is of the form \cite{ste95}
\begin{equation}\label{eq:SumSquaredLorentzians}
  S_{11}(\nu) =1+ \sum\limits_j \frac{\tilde A_j}{\nu^2 - \nu_j^2}
\end{equation}
where the $\nu_j$ are the complex valued resonance positions. As for all the resonances that will be studied in the sequel, one has for $|\textup{Re}(\nu_j)| \gg |\textup{Im}(\nu_j)|$ the approximation
\begin{equation}\label{eq:ApproxLorentzians}
  \frac{\tilde A_j}{\nu^2 - \nu_j^2} = \frac{\tilde A_j}{(\nu - \nu_j)(\nu + \nu_j)} \approx \frac{A_j}{\nu - \nu_j},
\end{equation}
where $A_j= \frac{\tilde A_j}{2\nu_j}$ is good for $\textup{Re}(\nu) >0 $. This leads to
\begin{equation}\label{eq:SumLorentzians}
  S_{11}(\nu) =1+ \sum\limits_j \frac{A_j}{\nu - \nu_j},
\end{equation}
which is more convenient in the following. In Fig.~\ref{fig:Spec} the spectra for three different $R/a$ parameters are shown. For the closed system ($R/a$=2, black solid line) even in the high frequency regime some separate resonances are visible. For $R/a$=2.25 the opening between the half circle and the straight wall is approximately 24\,mm. Thus, the opening supports approximately four modes in the shown frequency range and the resonances can still be recognized but are sufficiently broadened (red dotted line). By further increasing the opening ($R/a$=3.83, blue dashed line) the resonances are strongly overlapping.

In order to extract the resonances from this frequency signal we use the Harmonic Inversion (HI) algorithm as it has been presented by Main et\,al.~\cite{mai00}. To illustrate the challenges of applying this algorithm to experimental data, we first give a short summary of the algorithm described in \cite{mai00} and refer to \cite{mai00,kuh08b} for further details. Afterwards we propose some new tools to circumvent the occurring problems.

The HI algorithm extracts the resonances from the signal in four steps: windowing, truncating the Fourier transformed signal, Pad\'{e} approximation, and finally filtering. In the first step the measured signal is divided into overlapping windows. This is necessary to reduce the number of resonances in each separately analyzed window, which is crucial for applying the Pad\'{e} approximation. In order to reduce boundary effects of the windowing, only resonances within a given range $\Delta\nu_r$ in the center of each of the windows will be taken into account. The windows must therefore overlap correspondingly. In the following $\Delta\nu_r$=1\,GHz with a buffer region of 1\,GHz on each side leading to a total window size of 3\,GHz.

\begin{figure}
\includegraphics[width= 0.45\textwidth]{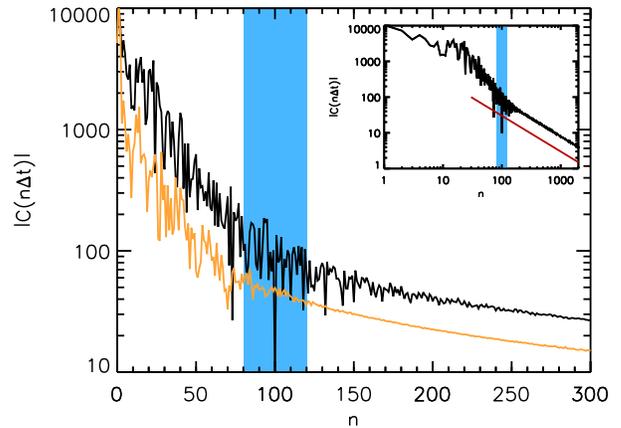}
\caption{\label{fig:FFT} (Color online)
Fast Fourier transform of the reflection signal of the frequency window from 20 to 23\,GHz for the closed system ($R/a =2$, upper black curve) and an open system ($R/a=3.83$, lower orange curve). The unit of the $x$ axis is the number of data points of this discrete time signal. The inset shows the decay for the closed system in a double logarithmic plot. The red straight line correspond to a $1/n$-decay. The light blue shaded region indicates the variation of the cutoff parameter between 80 and 120 data points, which is used in the data analysis described in Sec.~\ref{sec:ExpResults}.}
\end{figure}

For each of these windows a fast Fourier transform (FFT) is applied. Using (\ref{eq:SumLorentzians}) and neglecting the window effects, the FFT signal is given by
\begin{equation}\label{eq:Ct}
 C(t) = \delta(t)+\sum_j d_j e^{-2\pi i\nu_j t}.
\end{equation}
As the system is open, the resonance frequencies have a negative imaginary part and $C(t)$ decays exponentially with $t$. In Fig.~\ref{fig:FFT} examples for two different openings are shown. The decay at the beginning given by a sum of exponentials is due to the widths of the resonances and then changes to a $1/n$ decay which is due to the windowing, which can be seen in the inset. Applying the FFT to a measured window of 3\,GHz width and a step width of 0.1\,MHz we obtain a discrete time series of 30\,000 data points, which is still too large to apply the Pad\'{e} approximation. This problem is solved by truncating the time signal after $N_\mathrm{trunc}$ data points, which is justified by the fact that for high $t$ values the signal is governed by the $1/t$ decay which is due to the discontinuities at the window boundaries. The physical information of the decay is hence contained in the first part of the time signal. As it will be crucial in the sequel that the analyzed time series is of form (\ref{eq:Ct}) it
is important that one cuts off the signal before the $1/t$ decay takes over. On the other hand, $N_\mathrm{trunc}$ should also be chosen sufficiently large such that sufficient information is contained in the truncated signal. Figure~\ref{fig:FFT} shows that for the analyzed microwave spectra and a total window size of 3\,GHz a truncation at $N_\mathrm{trunc} = 80\dots 120$ is reasonable.

After this truncation an equidistant discrete time series of $N_\mathrm{trunc}$ data points $C(\Delta t\cdot n) = \sum_j d_j e^{-2\pi i\nu_j \Delta t\cdot n}$ with $n=0\dots N_\mathrm{trunc}-1$ remains and the central step is now to interpret this signal as a system of nonlinear equations of the form
\begin{equation}\label{eq:EqSys}
  c_n = \sum_{k=1}^{N_\mathrm{trunc}/2} d_k z_k^n \quad \textrm{ with } n=1\dots N_\mathrm{trunc}/2
\end{equation}
which is solved by the Pad\'{e} approximation \cite{mai00, pad92}.

\begin{figure}
\includegraphics[width= 0.45\textwidth]{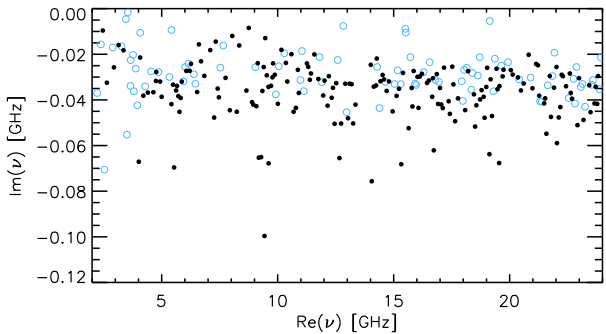}\\
\caption{\label{fig:Filter} (Color online)
This figure shows the resonances of the five disk system with $R/a=3.83$ in the complex plane obtained via the HI for one set of parameters. The light blue circles are resonances that are removed by the filter, whereas the black dots correspond to resonances that are assumed to be true.}
\end{figure}

An obvious problem with this approach is that $N_\mathrm{trunc}$ determines the number of resonances which are returned. Choosing this parameter according to the criteria discussed above $\frac{N_\mathrm{trunc}}{2}$ will be larger than the true number of resonances and the Pad\'{e} approximation will generate spurious resonances which have to be filtered out. There are two filter mechanisms: (i) The Pad\'{e} approximation is performed a second time with a time signal shifted by one point, that is, $C(\Delta t\cdot n)$ with $n = 1\dots N_\mathrm{trunc}$. Resonances that are unstable with respect to this point shift are rejected. (ii) Only resonances whose heights $\frac{|A_j|}{\left|\mathrm{Im}{\nu_j}\right|}$ exceed sufficiently the noise level are accepted. For the data analyzed in this article it has proven to be convenient to set this filter to 0.05. The effect of the filtering is illustrated in Fig.~\ref{fig:Filter}. Without filtering, the HI returns too many resonances; especially there appear resonances
with very small widths which are not realistic for such an open system. After filtering, only the resonances shown as black dots survive and the other resonances (light blue circles) are sorted out. For the remaining resonances a spectral gap and a conglomeration around $\mathrm{Im}(\nu) \approx -0.03$ can be observed. Have in mind that the imaginary part includes terms coming from the absorption in the top, bottom, and side walls and from the antenna. The description of the algorithm above shows that the HI algorithm requires the choice of several parameters, such as $N_\mathrm{trunc}$ or the window overlap. For all these parameters there is a range of plausible values; their exact values, however, are arbitrary. The assumption that the time series is a superposition of decaying exponentials only is another problem. For experimental data this will never be exactly true due to inevitable experimental noise and errors.

\begin{figure}
\includegraphics[width= 0.45\textwidth]{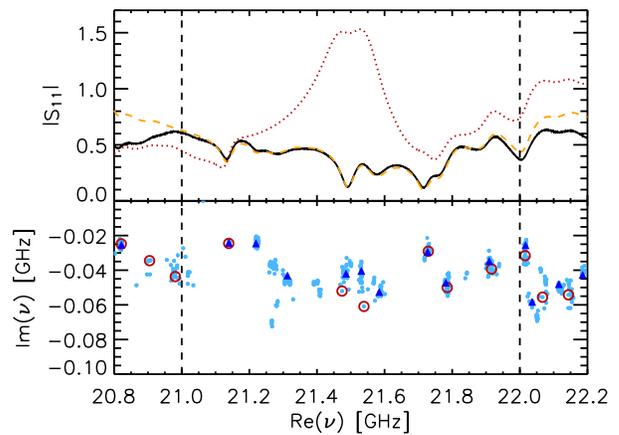}\\
\caption{\label{fig:BeWoRec} (Color online)
The top plot shows the measured microwave spectrum ($R/a = 3.83$) (black solid line) as well as two reconstructions obtained by the HI results for two different parameter sets (orange dashed and red dotted lines). The bottom plot shows the corresponding complex resonance positions. The red circles correspond to the resonances of the reconstructed signal shown above as red dotted line and the blue triangles to the orange dashed line. Light blue dots represent the resonance positions of the HI results for all parameter sets. The two vertical dashed lines mark the valid range of the analyzed window.}
\end{figure}

The possible problems occurring with the harmonic inversion of experimental data are visualized in Fig.~\ref{fig:BeWoRec}. The measured microwave spectrum for $R/a=3.83$ (black solid line) was analyzed by the harmonic inversion in order to extract the resonances in the valid range between 21 and 22\,GHz. Therefore, the HI was applied to the data with 160 different parameter sets by varying $N_\mathrm{trunc}$ between 80 and 120 and changing the buffer region on both sides between 1 and 0.98\,GHz, respectively. According to Fig.~\ref{fig:FFT} all these parameter sets are plausible. If the complex resonance positions of these 160 HI results are plotted in one plot (light blue dots in the lower part of Fig.~\ref{fig:BeWoRec}) one observes that the resonances of the different parameter sets form clusters in the complex plane. If one looks, however, at an individual result of a particular parameter set marked by the red circles one observes, that for this parameter set several clusters are missing. For other
parameter sets, however, nearly all the clusters within the valid interval are matched in the resulting resonances. The blue triangles mark one example.

Useful to reject bad HI results is the quality of the reconstruction which is calculated as follows: Suppose one is interested in resonances in the frequency interval from 21 to 22\,GHz as in Fig.~\ref{fig:BeWoRec} and has chosen to take the overlap on each side to be 1\,GHz. The HI algorithm thus analyzes the whole interval from 20 to 23\,GHz, Fourier transforms it, applies the Pad\'{e} approximation, and filters the resonances as described above. As a result it returns a set of resonances and amplitudes $\{(A_j, \nu_j)\}$, where the real parts of the found resonances range from 20 to 23 GHz. One now compares the superposition $\sum_j \frac{A_j}{\nu - \nu_j}$ with the measured signal $S_{11}(\nu)$ between 21 and 22\,GHz. Note that all the resonances in the large window range between 20 and 23 GHz are already contained in the superposition. However, influences of resonances outside this window and very broad resonances which cannot be extracted from the signal are not yet taken into account. We suppose that
these influences can be approximated over the size of the valid region by a complex valued linear function. This background is thus fitted to the difference between superposition and measured signal. The result of the reconstruction (which now contains the superposition of Lorentzians and the fitted background) is plotted in the top part of Fig.~\ref{fig:BeWoRec}. While the reconstructed signal (orange dashed line) of one set of resonances (blue triangles) agrees well with the measured data, the deviation of another reconstruction (red dotted line) using the other set (red circles) is enormous.

Based on these observations it is not sufficient to use only one parameter set for applying the HI on such kind of experimental data, as even the set leading to the best reconstruction might overlook single resonances. Therefore, we perform the HI for several parameter sets, reject those with unconvincing reconstruction by means of their $\chi^2$ value, and average over all the others.


\section{Counting Resonances: Experimental Results}
\label{sec:ExpResults}

In order to study the spectral asymptotics of the five-disk system, we have to determine the counting function (\ref{eq:DefCounting}) from the measured reflection spectrum. In this section we first give the detailed parameters used for the data analysis. Afterwards we present the results on the spectral asymptotics.

The resonance positions which are needed for the calculation of the counting function are extracted from the reflection spectra by the harmonic inversion. Thus according to Sec.~\ref{sec:HI} we start the analysis by decomposing the complete signal ranging from 1 to 25\,GHz into smaller windows. The width of the valid interval is chosen to be 1\,GHz and the buffer region on both sides as 1\,GHz as well. Thus, we obtain in total 22 windows allowing us to extract the resonances between 2 and 24\,GHz. For each of these windows we applied the HI with various different parameter sets for the reasons given in the preceding section. We varied $N_\mathrm{trunc}$ between 80 and 120, the shaded area in Fig.~\ref{fig:FFT}. Additionally, we slightly shifted the buffer region on both sides by 0.02\,GHz. The combination of these variations thus yields 160 different parameter sets.

For each window and each parameter set the HI returned a set of complex resonances and amplitudes. For each of these results the reconstruction had been compared to the original signal as described above and all parameter sets leading to a reconstruction with a $\chi^2$ error worse than the variance of the signal had been rejected. Now the density of states in the valid range of each window is obtained by averaging over the results of all accepted parameter sets. The \emph{averaged} counting function finally was obtained by the averaged density of states of all windows.

\begin{figure}
\includegraphics[width=.8\columnwidth]{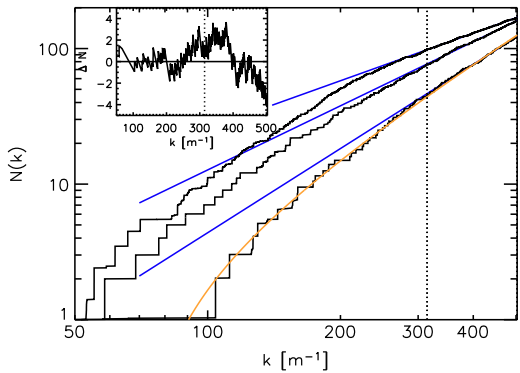}
\caption{\label{fig:density}(color online)
The counting functions for $R/a$=2, 2.25, and 3.9 (in this order from bottom to top) are plotted in black (histograms). Fits of their slope in the frequency range 15-24\,GHz, corresponding to a $k$ range of 315-500\,mm$^{-1}$ (dotted vertical line) are shown in blue (straight lines). The light orange curve over the lower histogram corresponds to the Weyl formula with 12\% loss for the closed system. Plotted in the inset is the difference between the Weyl formula with 12\% loss and the experimental counting function for the closed system ($R/a$=2).}
\end{figure}

In Fig.~\ref{fig:density} three examples of the experimentally obtained counting functions for different $R/a$-parameters are shown: for the closed ($R/a = 2$) and the most open system ($R/a = 3.9$) as well as for the transition region ($R/a = 2.25$). The closed curve is compared to the prediction of the Weyl law (\ref{eq:WeylClosed}) with area and boundary term. Measuring only with one fixed antenna we cannot expect to find all resonances as several of them have an amplitude of the order of the noise. In closed room temperature aluminum cavities of similar sizes this typically lead to a loss of 5\%-10\%. For $R/a=2$ the orange line shows the Weyl law (\ref{eq:WeylClosed}) with 12\% loss and is in good agreement with the extracted counting function. The inset shows the difference between Weyl law with 12\% loss and counting function. The experimental data fluctuates around the theory with a deviation of less than four resonances, but there is no overall tendency.

For the open systems the fractal Weyl law only predicts the asymptotic exponent of the counting function. To compare our results with this prediction the slope was fitted with a standard regression (blue lines) to our experimental counting functions in the interval 15-24\,GHz (marked by the dotted vertical line). For the fit we need, on the one hand, a sufficiently large range to extract the slope reliably; on the other hand, we need large $k$ to get into the semiclassical regime. In the chosen fit range all counting functions show an approximately linear behavior.

\begin{figure}
\includegraphics[width=.8\columnwidth]{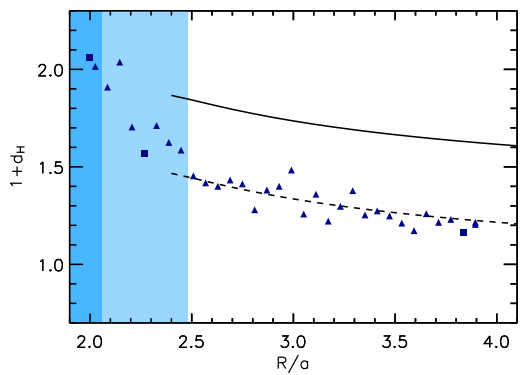}
\caption{\label{fig:exponent} (Color online)
The data points correspond to the fitted exponent of the counting function in dependence of the $R/a$ parameter. The solid line shows the asymptotic exponent $1+d_H$ predicted by the fractal Weyl law and the dashed curve is the same shifted by 0.4 as a guide for the eye. The three squares mark the examples which have already been presented in the previous figures. The darker shaded blue region indicates the $R/a$ values without open channels whereas in the range marked by the lighter shaded blue region only a few open channels (1, \dots, 8) exist.}
\end{figure}

In Fig.~\ref{fig:exponent} the slopes of the counting functions are plotted versus their $R/a$ parameter. For the quantum-mechanically closed system the classical Weyl law predicts a value of 2. As long as the slit between disk and metallic wall is smaller than half a wavelength of the maximal frequency 24\,GHz, the system is quantum mechanically closed and couples only by tunneling to the exterior. This region is highlighted by the darker blue shading and an exponent of 2 is still expected within the wave number range investigated. For large $R/a$ parameter the fractal Weyl law predicts a fractional exponent. The black solid line shows the predicted exponent $1+d_H$ of the fractal Weyl law, where $d_H$ is the reduced fractal dimension of the repeller. This dimension has been calculated via the topological pressure and cycle expansion in dependence of $R/a$ (see \cite{gas89b,gas89a,cvi89} for details). Due to pruning effects this method fails to calculate $d_H$ for small $R/a$ parameters; thus, the black theoretical line stops at $R/a=2.4$. In between there will be a transition region. In Fig.~\ref{fig:exponent} the region between one and eight open channels is shaded in light blue. Experimentally, the expected start value of about 2 for the closed system and a smooth transition to lower noninteger values between 1 and 2 is seen. For the quantum-mechanically closed system the fitted exponent agrees well with the predicted value of 2 of the classical Weyl law. In the transition region with a few open channels the exponent decreases smoothly. For the open system the fitted exponents do not match the theoretically predicted curve, but the parametric dependence still can be seen though the theoretical curve has to be shifted down by 0.4 in order to match the fitted exponents. However, the fitted exponents are significantly larger than one and definitely non integer.

A possible explanation for the down shifted fitted exponents could be the following. Suppose that the probability that the harmonic inversion overlooks some resonances, increases when the resonances become more and more overlapping. The density of states for the five-disk system is supposed to increase like $k^{d_H}$; thus, with increasing $k$ the resonances become stronger overlapping and the supposed loss of resonances increases as well. Such a  loss which increases with $k$ would, however, lead to a systematically lower fitted exponent. Even if the procedure of averaging over many HI parameter sets and checking the reconstruction significantly increases the reliability of the results, those ambiguities will remain. This process might also explain why for the closed systems this shift does not occur: As the resonance for the more closed systems are significantly less overlapping even for high frequencies (see Fig.~\ref{fig:Spec}) the loss mechanism might not yet have set in.

Moreover, there are, of course, the experimental uncertainties influencing the findings: The exponential behavior of the counting function is predicted in the semiclassical limit, corresponding to the limit of infinite frequencies. Experimentally, we are, however, restricted to a finite frequency and width range. Additionally, the resonance structure of such open microwave systems is very sensitive to inevitable reflections on imperfect absorbers.

\section{Conclusion}
\label{sec:Conclusion}

In this article we experimentally examined the behavior of the counting function in a five-disk microwave system which was successively transformed from a closed to an open system with a classically fractal repeller. The counting function was obtained from the measured spectrum by the harmonic inversion. We pointed out the occurring problems and proposed some tools to circumvent them. In the closed regime we found good agreement between the extracted counting function and theory. In the range of many open modes the functional dependence showed an agreement with theoretical predictions, though the experimental exponents had been found to be approximately 0.4 below the prediction. This certainly is in agreement with mathematics upper bounds though falls short of demonstrating asymptotics (\ref{eq:FractalWeyl}). It is, however, clear that the growth of resonances is not linear as predicted in \cite{gas89b}. In between there seems to exist a smooth monotonic transition between the two regimes, exhibiting the existence of a small number of open channels for the investigated wave-number range.


\begin{acknowledgments}
We are thankful for St\'{e}phane Nonnenmacher and Bruno Eckhardt for intensive discussions. This work was supported by the Deutsche Forschungsgemeinschaft via an individual grant and the Forschergruppe 760 `Scattering systems with complex dynamics'. TW acknowledges financial support of the `German National Academic Foundation'.
\end{acknowledgments}

\end{document}